# Discovery of new boron-rich chalcogenides: orthorhombic $B_6X$ (X = S, Se)

K.A. Cherednichenko,[1] V.A. Mukhanov,[1] Z. Wang,[2,3] A.R. Oganov,[2,4,5] A. Kalinko,[6,7]
I. Dovgaliuk,[8]  and  V.L. Solozhenko [1,*]

[1] *LSPM–CNRS, Université Paris Nord, 93430 Villetaneuse, France*

[2] *Skolkovo Institute of Science and Technology, Skolkovo Moscow Region, 143026, Russia*

[3] *Nanjing University of Posts and Telecommunications, Nanjing, Jiangsu 210003, China.*

[4] *Moscow Institute of Physics and Technology, Dolgoprudny City, Moscow Region 141700, Russia.*

[5] *School of Materials Science, Northwestern Polytechnical University, Xi'an 710072, China.*

[6] *Institute of Solid State Physics, University of Latvia, LV-1063 Riga, Latvia*

[7] *Universität Paderborn, Naturwissenschaftliche Fakultät, 33098 Paderborn, Germany*

[8] *European Synchrotron Radiation Facility, 38043 Grenoble, France*

**Abstract**

New boron-rich sulfide $B_6S$ and selenide $B_6Se$ have been discovered from high pressure – high temperature synthesis combined with *ab initio* evolutionary crystal structure prediction, and studied by synchrotron X-ray diffraction and Raman spectroscopy at ambient conditions. As it follows from Rietveld refinement of powder X-ray diffraction data, both chalcogenides have orthorhombic symmetry and belongs to *Pmna* space group. All experimentally observed Raman bands have been attributed to the theoretically calculated phonon modes, and the mode assignment has been performed. Prediction of mechanical properties (hardness and elastic moduli) of new boron-rich chalcogenides have been made using *ab initio* routines, and both compounds were found to be members of a family of hard phases ($H_V \sim 31$ GPa).

## I. Introduction

Development of modern industry requires more new materials with exceptional physical and chemical properties. Searching for such materials becomes a central challenge of modern materials science. The discoveries of fullerene, carbon nanotubes and graphene unveiled that unusual crystal structures open access to the unique properties.

Boron-rich compounds are the materials possessing such unusual structures. The $B_{12}$ *closo*-clusters are the common feature of these compounds. Almost all boron-rich solids may be considered as a combination of 'electron deficient' $B_{12}$-icosahedral units (36 valence electrons over 48 bonding orbitals) and various interstitial atoms (from nonmetals to metals) as electron donors [1],[2]. The polycentric metal-like bonding system within the $B_{12}$ icosahedra and strong covalent bonds between $B_{12}$ *closo*-clusters and interstitial atoms makes boron-rich compounds extremely stable, which hereby leads to high melting temperatures, chemical inertness, outstanding mechanical properties, etc. [1]-[4]. A change of the interstitial atoms makes possible to considerably vary the properties (e.g. bulk moduli variation of α-rhombohedral boron (α-$B_{12}$) and isostructural boron-rich compounds: $B_4C$, $B_{12}O_2$, $B_{13}N_2$, $B_{12}P_2$, etc. [5],[6]). Thus, a detailed study of already existing materials and exploration of new boron-rich compounds are of great importance and draw considerable attention in experiment and theory.

In the present work two new boron-rich chalcogenides were synthesized under extreme pressure–temperature conditions. Their crystal structures were found by *ab initio* crystal structure prediction, which allowed us to perform Rietveld refinement of the experimental X-ray diffraction (XRD) patterns. The Raman spectra of both boron-rich chalcogenides were acquired at ambient conditions, and the observed Raman bands were assigned to the corresponding phonon modes.

## II. Methods

### A. *Experimental*

Formation of new boron-rich chalcogenides was first observed in our *in situ* high pressure – high temperature studies of the B–S and B–Se binary systems at BL04B1 beamline, SPring-8 (Japan) and PSICHE beamline, SOLEIL (France). Chemical interaction of elemental boron with sulfur and selenium melts were studied by energy-dispersive X-ray diffraction at pressures up to 11 GPa and temperatures up to 2500 K using SPEED-1500 multianvil press (BL04B1) and Paris-Edinburgh press (PSICHE) using white beam (20-150 keV, bending magnet @ BL04B1; 25-80 keV, wiggler source @ PSICHE).

Based on the information about the most appropriate synthesis conditions and optimal stoichiometries of B:S(Se) reaction mixtures extracted from our synchrotron studies, the new boron-rich chalcogenides have been synthesized at 6.1 GPa and 2700 K in a toroid-type high-pressure apparatus. A design of the high-temperature assembly used in recovery experiments is described elsewhere [7]. The powders of amorphous boron (Grade I ABCR), and sulfur and selenium (both Alfa Aesar, 99.5%) were used as starting materials. Boron nitride capsules (COMPRES) were used to isolate the reaction mixture (B:X molar ratio 5:1) from the graphite heater. The recovered samples were treated with 3N nitric acid (ACS, Alfa Aesar) for 20 min at 370 K in order to remove unreacted boron, washed with deionized water and dried at 400 K.

X-ray diffraction study of boron-rich chalcogenides was performed at Swiss-Norwegian Beamline BM01, ESRF [8]. The wavelength of monochromatic beam from a bending magnet was set to 0.6866 Å. X-ray diffraction patterns were collected during 20 s in Debye-Scherrer geometry with rotating quartz-glass capillary using PILATUS 2M detector. The crystal structure refinement was performed using Maud software [9]; high purity $LaB_6$ was used as a standard.

Raman spectra of polished well-sintered polycrystalline samples were measured at ambient conditions in the 100-2000 cm$^{-1}$ range using Horiba Jobin Yvon HR800 Raman spectrometer. Unpolarized light from 488-nm line of an Ar laser (10 μm beam spot) was used for excitation. The spectrometer was calibrated using single-crystal cubic Si at room temperature.

## B. Computational Details

X-ray diffraction patterns of the newly synthesized phases clearly did not match any previously known phases. Neither their structures, nor the exact chemical compositions were known. Taking into account the starting B:S:Se molar ratios we assumed the probable composition as: $B_xS$ and $B_xSe$ with $5 \leq x \leq 7$. This information was insufficient for the determination of the crystal structures solely from experiment.

We performed variable-composition searches for all stable compounds in the B-S and B-Se systems using the USPEX code [10]-[12], which has already demonstrated exceptional predictive power, reliability and efficiency for discovering novel compounds and their crystal structures (e.g. [13]-[16]). Searches were performed at the pressure of 20 GPa, the initial population was made of structures containing up to 30 atoms in the primitive unit cell. In each generation there were 60 structures, and calculations were run for 60 generations. All produced structures were carefully relaxed and their enthalpies were computed using the Vienna *ab initio* Simulation Package (VASP) [17] within the generalized gradient approximation (GGA) of Perdew-Burke-Ernzerhof (PBE) [18]. Total energy was calculated within the framework of projector augmented wave (PAW) method [19]. We used plane wave energy cutoff of 550 eV and Gamma-centered $K$-point mesh with the resolution of $2\pi \cdot 0.06 Å^{-1}$ for final structural relaxations in USPEX. For mechanical and electronic property calculations, we improved the $K$-point mesh to the resolution of $2\pi \cdot 0.04 Å^{-1}$. We found that at 20 GPa the following boron sulfides and selenides are thermodynamically stable (see thermodynamic convex hulls in Fig. 1(*a*) and Fig. 1(*b*)): $B_6S$, BS, $B_2S_3$ for boron sulfides and $B_6Se$ and BSe for boron selenides. Nevertheless, $B_6S$ and $B_6Se$ were the only stable compounds satisfying our initial assumption about the chemical compositions of the synthesized phases. The computed enthalpies of the lowest-enthalpy structures as a function of pressure are shown in Fig. 1(*c*) and Fig. 1(*d*): $B_6S$ with *Pmna* space group is stable in 0-20 GPa pressure range, whereas the structure of $B_6Se$ with *Pmna* space group is stable in 4-20 GPa range. This structure matched experimental XRD patterns and provided a good starting model for Rietveld refinement.

The Raman spectra of both boron-rich chalcogenides were computed using VASP code with the fully relaxed structure. Firstly, we performed phonon calculation to determine phonon frequencies and normal modes at the Γ-point based on density functional perturbation theory (DFPT) as implemented in the PHONOPY code. Further DFPT method was used to compute out macroscopic dielectric tensor. And lastly, Raman intensity for each normal mode was obtained by calculating the derivative of the calculated macroscopic dielectric tensor (or polarizability) with respect to the corresponding normal mode coordinate.

At the same time, structural and phonon properties of both boron-rich chalcogenides were also studied using linear combination of atomic orbital (LCAO) calculations based on the hybrid exchange-correlation density functional (DFT)/Hartree-Fock (HF) scheme, which is implemented in CRYSTAL17 code [20]. For boron and sulfur atoms we used all electron basis sets which were optimized in earlier calculations [21],[22][20]. The core electrons of the selenium atoms were excluded from consideration using the effective core pseudopotential (ECP) with corresponding atomic basis set [22]. The accuracy of the calculation of the bielectronic Coulomb and exchange series is controlled by the set of tolerances, which were taken to be $10^{-7}$, $10^{-7}$, $10^{-7}$, $10^{-9}$, and $10^{-30}$, according to the recommendation for hybrid functionals [22]. The Monkhorst-Pack scheme [23] for an 8×8×8 k-point mesh in the Brillouin zone was applied. Self-consistent field calculations were performed for hybrid DFT/HF WCGGA-PBE-16% functional [24]. The percentage 16% defines the Hartree-Fock admixture in the exchange part of DFT functional.

The full structure optimization procedure according to the energy minima criterion was performed for both boron-rich chalcogenides (Table I and Table S1). The bulk moduli of both compounds were estimated using routine implemented in CRYSTAL17 code [25]. The unit cell volumes were varied from 95% to 105% of the volume ($V_0$) corresponding to the energy minimum ($E_0$). The structure optimization was performed at each volume value. The obtained E(V) dependences were fitted to the Birch-Murnaghan equation of state.

The phonon frequencies for both compounds were calculated using the direct (frozen-phonon) method implemented in CRYSTAL17 code [26],[27] (Table S2). Calculation of Raman intensities was performed by using a coupled-perturbed Hartree–Fock/Kohn–Sham approach [27],[28]. Raman spectra were constructed by using the transverse optical (TO) modes and by adopting a pseudo-Voigt functional form [26] with a full width half maximum parameter set to 1. The choice of the broadening was determined according to the criteria to keep maximal possible small intensity bands in theoretical spectrum, which are smeared out while applying higher broadening parameters. A visualization of the calculated phonon modes was performed using MOLDRAW software [29].

## III.  Results and Discussion

### A.  *Crystal structure of new boron-rich sulfide and selenide*

As mentioned above, the crystal structures of new boron-rich sulfide and selenide were predicted using the USPEX algorithm: both have the same orthorhombic symmetry and belong to the same space group *Pmna* (53). The predicted lattice parameters and mechanical properties as well as atoms positions are collected in Table I and Table S1, respectively.

Theoretically predicted crystal structures of boron-rich chalcogenides were further used as starting models in the Rietveld refinement of the powder X-ray diffraction patterns taken at ambient conditions (Fig. 2). The backgrounds of both diffraction patterns were approximated by a 5-order polynomial. The final reliability factors $R_{wp}$ converged to 5.0 (Fig. 2*a*) and 5.8 (Fig 2*b*) indicate the excellent refinement level and, thus, confirming the correctness of the structures found with USPEX algorithm. The refined lattice parameters of boron-rich sulfide and selenide are also presented in Table I.

The unit cell of both boron-rich chalcogenides contains 24 boron atoms in four independent (*4h* and *8i*) Wyckoff positions and 4 sulfur/selenium atoms placed in one independent (*4h*) Wyckoff position. Since all boron atoms constitute $B_{12}$ clusters their total atom site occupancies were fixed to 1.0 by default. The total S1 and Se1 sites occupancies were found to be 0.925 and 0.952, respectively. The details of atomic structure of both compounds are presented in Table S1. Considering the occupancies of S1 and Se1 sites are close to 1, the stoichiometry of new orthorhombic boron-rich sulfide and selenide may be presented as "$B_6X$", where X is S or Se. It should be underlined that the attempt to replace S and Se atoms by B atoms resulted in a large mismatching and high $R_{wp}$ values. For convenience and in order to avoid any further confusion with previously reported hexagonal boron-rich chalcogenides (e.g. $B_{12}S_{2-x}$ [30],[31] and $B_{12}Se_{(2-x)}B_x$ [32]) further we will call the new boron-rich sulfide and selenide as "$o$-$B_6S$" and "$o$-$B_6Se$" (where "$o$" indicates the orthorhombic symmetry). The unit cell of $o$-$B_6X$ (where X = S, Se) is presented in Fig. 3. The X-ray densities of $o$-$B_6S$ and $o$-$B_6Se$ were found to be 2.54 g/cm$^3$ and 3.58 g/cm$^3$, respectively which is in good agreement with values predicted *ab initio* using USPEX (2.53 g/cm$^3$ and 3.55 g/cm$^3$) and CRYSTAL17 (2.58 g/cm$^3$ and 3.66 g/cm$^3$).

Among all experimentally obtained nonmetal boron-rich compounds only $o$-$B_6S$ and $o$-$B_6Se$ have the orthorhombic structure (with exception of $B_6Si$ (*Pnnm*) [33] and $B_3Si$ (*Imma*) [34]). The distribution/packing of the $B_{12}$ *closo*-clusters in $o$-$B_6X$ (X = S, Se) unit cells may be described as side-centered (Fig. 3) similar to that in $B_3Si$. One slightly distorted $B_{12}$-icosahedron in $o$-$B_6S$ and $o$-$B_6Se$ is linked with six others. The lengths of intra-icosahedral B–B bonds vary from 1.7294 Å to 1.8987 Å in $o$-$B_6S$ and from 1.7077 Å to 1.9009 Å in $o$-$B_6Se$, whereas the inter-icosahedral bond lengths in $o$-$B_6S$ and $o$-$B_6Se$ are: 1.6949 Å (B1–B1), 1.7448 Å (B2–B2), and 1.7511 Å (B1–B1), 1.8004 Å (B2-B2) respectively. One sulfur/selenium atom is linked with three closest icosahedra: S1–B4 (1.8884 Å), S1–B3 (1.8586 Å), Se1–B3 (1.9623 Å) and Se1–B4 (2.0128 Å).

According to the theoretically predicted mechanical properties presented in Table I (the $B_0$ values estimated by VASP and CRYSTAL17 are in good agreement), the both boron-rich chalcogenides are considerably more compressible and less hard than γ-$B_{28}$ [35],[36], α-rhombohedral boron [37]-[39] and isostructural boron-rich compounds [5],[6],[40]-[44].

## B. Raman spectra of new boron-rich sulfide and selenide

$o$-$B_6S$ and $o$-$B_6Se$ have 28 atoms in the unit cell, thus, 84 normal modes are expected. According to the symmetry analysis, the acoustic and optic modes of $o$-$B_6X$ (where X = S or Se) at Γ point can be presented as follows:

$$\Gamma_{acoustic} = B_{1u} + B_{2u} + B_{3u}$$

$$\Gamma_{optic} = 12A_g + 9A_u + 9B_{1g} + 11B_{1u} + 9B_{2g} + 11B_{2u} + 12B_{3g} + 8B_{3u}$$

$11B_{1u} + 11B_{2u} + 8B_{3u}$ are IR active modes; $12A_g + 9B_{1g} + 9B_{2g} + 12B_{3g}$ are Raman active modes; others are dumb modes.

Raman spectra of $o$-$B_6S$ and $o$-$B_6Se$ were measured in the 100-2500 $cm^{-1}$ frequency range, however, all bands were observed in the 150-1100 $cm^{-1}$ region (Fig. 4). The Raman spectra of $o$-$B_6S$ and $o$-$B_6Se$ resemble the Raman spectra of α-$B_{12}$ [45],[46] and γ-$B_{28}$ [47] and other boron-rich compounds [48]-[51]. The most intense and narrow bands are situated in the low-frequency region (<500 $cm^{-1}$ for $o$-$B_6S$ and <400 $cm^{-1}$ for $o$-$B_6Se$), whereas the less intense and broad bands and band groups are concentrated in the high-frequency region.

The CRYSTAL17 and VASP calculated Raman spectra of both compounds (at T = 0 K) are presented in Fig. 4 and Fig. S1, respectively. The theoretically predicted Raman active phonon modes ($\omega_r^C$ and $\omega_r^V$, for CRYSTAL17 and VASP, respectively), experimentally observed Raman bands and overlapped band groups ($\omega_0$) of $o$-$B_6S$ and $o$-$B_6Se$ are listed in Table S2. The theoretical and experimental data were found to be in a good agreement. The average error on individual modes being less than 1.5% for $o$-$B_6S$ (with a maximum error of 2.8%; mode at 780 $cm^{-1}$) and 1.2% for $o$-$B_6S$ (with a maximum error of 2.7%; mode at 594 $cm^{-1}$). A good agreement between theory and experiment (also observed in our previous Raman studies [52],[53]) gave us confidence in the predictive power of our *ab initio* calculations for the modes assignment (see Table S2).

Thanks to MOLDRAW software [29] visualizing the calculated phonon modes, each experimentally observed Raman band or bands group can be confidently associated with the corresponding atomic movements in $o$-$B_6S$ and $o$-$B_6Se$ unit cells. Taking into account that normal modes of boron-rich chalcogenides with such complicated structure may incorporate various simultaneous atomic movements, we distinguished the most distinct ones only for the convenience of the further modes description. As one can see in Fig. 4, the Raman bands of both spectra were divided onto four groups (*G1 − G4*).

The "*G1*" contains the low frequency modes (280-400 $cm^{-1}$ for $o$-$B_6S$ and 150-280 $cm^{-1}$ for $o$-$B_6Se$) referred to the symmetric and asymmetric oscillations (e.g. rocking, twisting, wagging) of S/Se atoms and the corresponding $B_{12}$–icosahedral units distortions.

The Raman bands of group "*G2*" (420-520 $cm^{-1}$ for $o$-$B_6S$ and 330-500 $cm^{-1}$ for $o$-$B_6Se$) were associated with various tilting oscillations of the whole $B_{12}$ units around different unit cell directions (rocking and wagging of the equatorial and polar boron atoms of one $B_{12}$-unit). Unlike Se atoms, the oscillations of S atoms were found rather significant in some "*G2*" modes. This phenomenon can be easily explained by the atomic mass difference of S and Se atoms.

The middle-frequency modes in "*G3*" (550-760 $cm^{-1}$ for $o$-$B_6S$ and 550-740 $cm^{-1}$ for $o$-$B_6Se$) were referred, first of all, to different vibrations of the equatorial B atoms (B2 - B4) leading to stretching of the intra-icosahedral B–B bonds, rotations of the B1–B1 and B2–B2 inter-icosahedral bonds and rotations, twisting and "parasol" oscillations of the S–(B)$_3$ structural elements (three B atoms belong to three different B$_{12}$-units).

The "*G4*" group contains the high-frequency modes (760-1100 cm$^{-1}$ for *o*-B$_6$S and 740-1050 cm$^{-1}$ for *o*-B$_6$Se) described by oscillations of the equatorial and polar born atoms of B$_{12}$ units leading to stretching of the inter-icosahedral bonds (B–Se, B2–B2, B1–B1). For instance, in both spectra the two phonon modes with the highest frequencies correspond to the oscillations of the polar B1 atoms and, thus, to the stretching of the B1–B1 inter-icosahedral bonds.

Such a division of *o*-B$_6$S and *o*-B$_6$Se normal modes is consistent with previously reported classification of vibrational modes of α-B$_{12}$ and isostructural boron-rich compounds [45],[46],[49],[50]: the modes involving the whole icosahedron rotations lay in the 100-200 cm$^{-1}$ range, intra-icosahedral modes lay between 550-950 cm$^{-1}$, and inter-icosahedral modes with wave numbers above 1000 cm$^{-1}$.

The detailed explanation of the bands widths over ~600 cm$^{-1}$ requires additional XRD and Raman studies of *o*-B$_6$S and *o*-B$_6$Se single crystals (perhaps coupled with low-temperature and high-pressure measurements). Nonetheless, it might be assumed, that some random distortions of B$_{12}$-icosahedral units (not detectable by powder XRD) and, thus, corresponding distortion of the intra- and inter-icosahedral bonds as well as partial occupation of *4h* sites by S/Se atoms might be the most probable reasons of the observed Raman bands broadening. Earlier, the isotopic $^{11}$B/$^{10}$B disorder in α-boron was also proposed as a possible reason of the Raman bands broadening [46].

## IV.   Conclusions

New boron-rich sulfide *o*-B$_6$S and selenide *o*-B$_6$Se were synthesized under extreme *p-T* conditions and studied by powder X-ray diffraction and Raman spectroscopy at ambient pressure. According to *ab initio* evolutionary crystal structure predictions combined with Rietveld refinement of synchrotron X-ray diffraction data, both boron-rich chalcogenides have orthorhombic symmetry and belong to *Pmna* (53) space group. The observed Raman bands were assigned to the phonon modes and associated with atomic movements. Elastic properties of new boron-rich chalcogenides have been theoretically predicted using various *ab initio* routines.

### Acknowledgements

The authors thank T. Chauveau (LSPM) for help with Rietveld analysis, and Drs. Y. Tange (SPring-8) and N. Guignot (SOLEIL) for assistance in synchrotron experiments that were carried out during beamtimes allocated to proposals 2017A1047 & 2018A1121 at SPring-8 and proposal 20170092 at SOLEIL. *Ab initio* calculations have been performed using Rurik and Arkuda supercomputers.

This work was financially supported by the European Union's Horizon 2020 Research and Innovation Programme under Flintstone2020 project (grant agreement No. 689279). Z.W. thanks the National Science Foundation of China (grant No. 11604159). A.R.O. thanks the Russian Science Foundation (grant 16-13-10459).

**Table I.** Unit cell parameters ($a_0$, $b_0$, $c_0$) and predicted mechanical properties of $o$-B$_6$X (X = S, Se): bulk modulus ($B_0$), shear modulus ($G$), Young's modulus ($E$) and Vickers hardness ($H_V$).

|  | $o$-B$_6$S (*Pmna*) | | | $o$-B$_6$Se (*Pmna*) | | |
|---|---|---|---|---|---|---|
|  | Exp. | VASP | CRYSTAL17 | Exp. | VASP | CRYSTAL17 |
| $a_0$, Å | 5.8170(1) | 5.8307 | 5.8139 | 5.9463(1) | 5.9684 | 5.9359 |
| $b_0$, Å | 5.3025(1) | 5.3202 | 5.2918 | 5.3579(1) | 5.3802 | 5.3416 |
| $c_0$, Å | 8.2135(1) | 8.2072 | 8.2026 | 8.3824(1) | 8.3809 | 8.3631 |
| V$_0$, Å$^3$ | 253.34(1) | 254.59 | 252.36 | 267.06(1) | 269.12 | 265.17 |
| $B_0$, GPa | — | 146 | 151 | — | 138 | 144 |
| $G$, GPa | — | 138 | — | — | 134 | — |
| $E$, GPa | — | 315 | — | — | 304 | — |
| $H_V$, GPa | — | 31 | — | — | 31 | — |

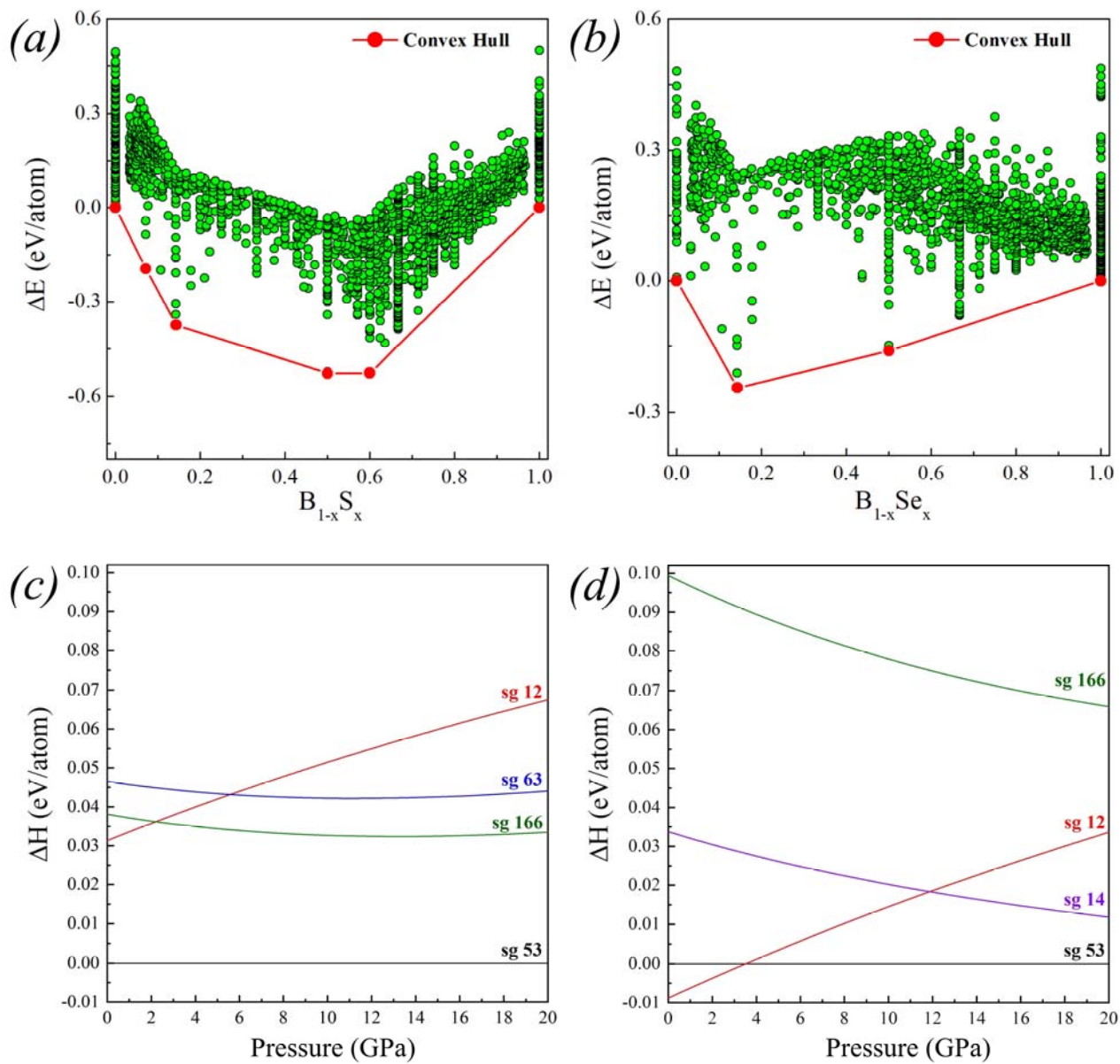

**Figure 1** Convex hull of B-S (*a*) and B-Se (*b*) variable-composition USPEX calculations at 20 GPa. Enthalpy difference ($\Delta$H) between stable/metastable $B_6S$ (*c*) and $B_6Se$ (*d*) structures (*sg* is space group) in the 0-20 GPa pressure range.

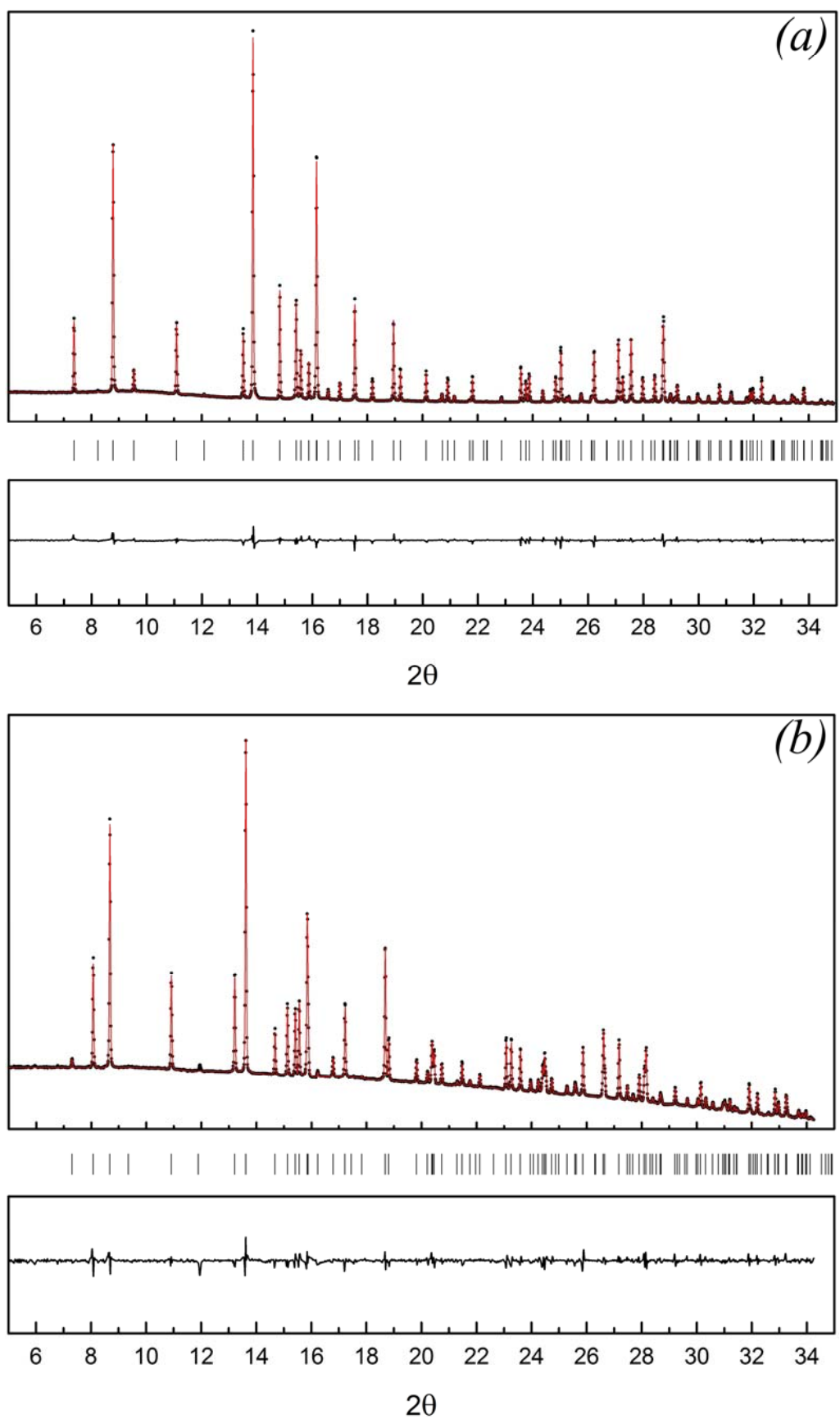

**Figure 2** Rietveld full profile refinement of powder X-ray diffraction patterns of *o*-B$_6$S (*a*) and *o*-B$_6$Se (*b*).

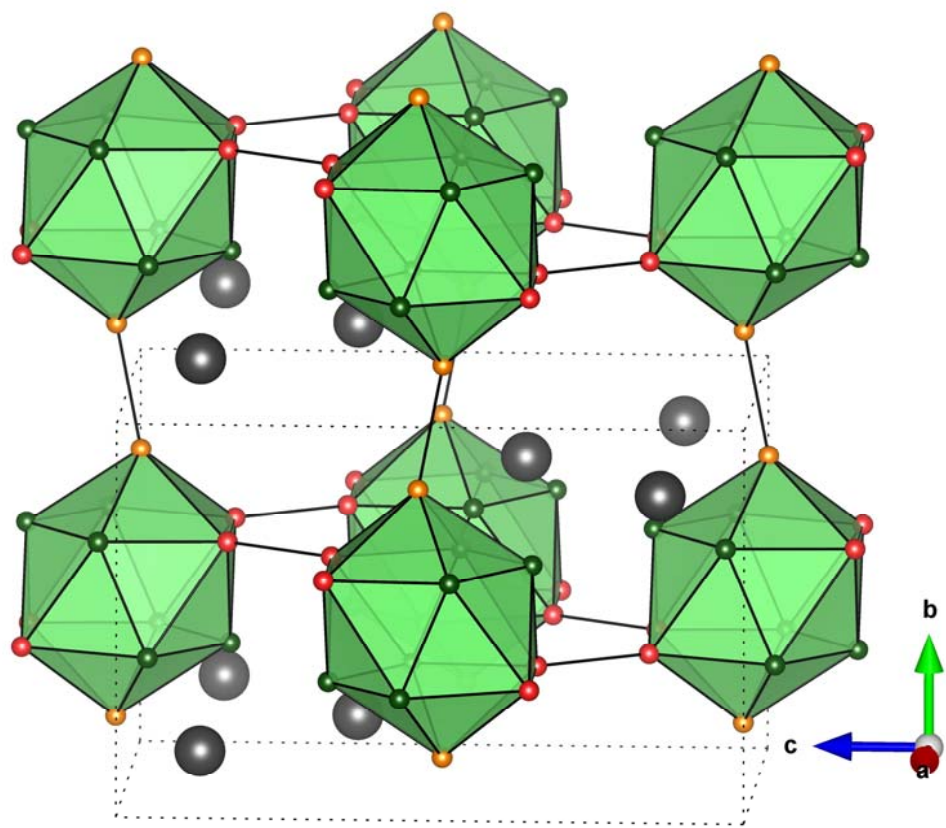

**Figure 3**   Crystal structure of new orthorhombic phases of boron-rich sulfide and selenide, $o$-B$_6$X (B$_{12}$-units are presented by green icosahedral polyhedral; polar B1; equatorial B2 and B3; and B4 atoms are marked by orange, red and green balls, respectively; X = S, Se atoms are shown as large grey balls).

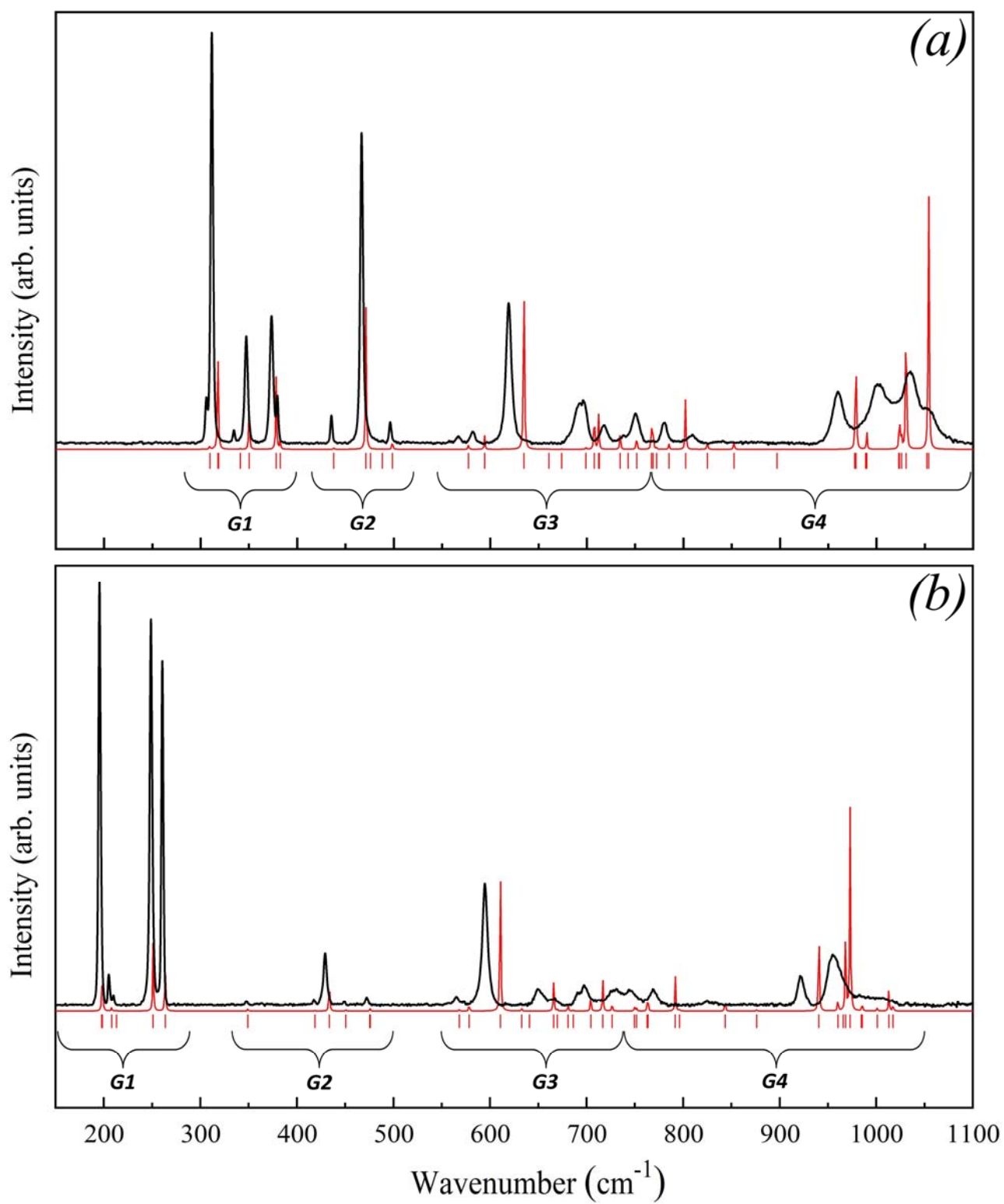

**Figure 4**    The experimental (black) and calculated by CRYSTAL17 (red) Raman spectra of *o*-B₆S (*a*) and *o*-B₆Se (*b*). The red dashes show all predicted Raman active phonon modes.

Supporting Information

# Discovery of new boron-rich chalcogenides: orthorhombic $B_6X$ (X = S, Se)

K.A. Cherednichenko, V.A. Mukhanov, Z. Wang, A.R. Oganov, A. Kalinko, I. Dovgaliuk and V.L. Solozhenko*

**Abstract:** New boron-rich sulfide $B_6S$ and selenide $B_6Se$ have been discovered from high pressure – high temperature synthesis combined with *ab initio* evolutionary crystal structure prediction, and studied by synchrotron X-ray diffraction and Raman spectroscopy at ambient conditions. As it follows from Rietveld refinement of powder X-ray diffraction data, both chalcogenides have orthorhombic symmetry and belongs to *Pmna* space group. All experimentally observed Raman bands have been attributed to the theoretically calculated phonon modes, and the mode assignment has been performed. Prediction of mechanical properties (hardness and elastic moduli) of new boron-rich chalcogenides have been made using *ab initio* routines, and both compounds were found to be members of a family of hard phases ($H_V$ ~ 31 GPa).

**Results and Discussion**

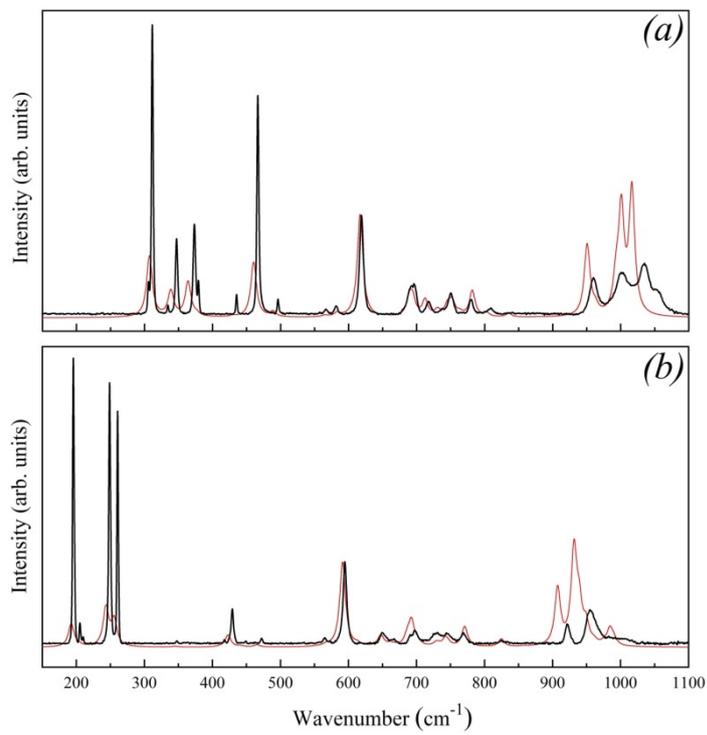

**Figure S1.** The experimental (red) and VASP-based calculated (black) Raman spectra of $o$-B$_6$S ($a$) and $o$-B$_6$Se ($b$) at ambient conditions.

**Table S1.** The atomic structure (first predicted by USPEX) of orthorhombic boron-rich sulfide and selenide retrieved from Rietveld refinement and *ab initio* geometry optimizations (VASP and CRYSTAL17).

| | Phase | Atom label (Wyckoff) | $x$ | $y$ | $z$ | $B_{iso}$, Å$^2$ | Site occupancy |
|---|---|---|---|---|---|---|---|
| *Experimental* | *o*-B$_6$S | S1 (*4h*) | 0.0000 | 0.1863(1) | 0.8681(1) | 1.401(3) | 0.925(3) |
| | | B1 (*4h*) | 0.0000 | 0.1578(6) | 0.4835(4) | 2.05(8) | 1.0[f] |
| | | B2 (*8i*) | 0.8385(4) | 0.3565(4) | 0.3358(3) | 2.23(6) | 1.0[f] |
| | | B3 (*4h*) | 0.0000 | 0.3454(6) | 0.6665(5) | 2.39(8) | 1.0[f] |
| | | B4 (*8i*) | 0.2591(5) | 0.3511(5) | -0.0432(3) | 2.34(8) | 1.0[f] |
| | *o*-B$_6$Se | Se1 (*4h*) | 0.0000 | 0.1695(1) | 0.8663(9) | 0.31(2) | 0.952(5) |
| | | B1 (*4h*) | 0.0000 | 0.1605(12) | 0.4803(7) | 1.45(18) | 1.0[f] |
| | | B2 (*8i*) | 0.8402(5) | 0.3638(8) | 0.3363(6) | 0.83(14) | 1.0[f] |
| | | B3 (*4h*) | 0.0000 | 0.3493(12) | 0.6623(8) | 0.39(14) | 1.0[f] |
| | | B4 (*8i*) | 0.2638(7) | 0.3531(11) | -0.0396(5) | 0.58(13) | 1.0[f] |
| VASP | *o*-B$_6$S | S1 (*4h*) | 0.0000 | 0.1879 | 0.8698 | — | 1.0 |
| | | B1 (*4h*) | 0.0000 | 0.1631 | 0.4829 | — | 1.0 |
| | | B2 (*8i*) | 0.8417 | 0.3628 | 0.3361 | — | 1.0 |
| | | B3 (*4h*) | 0.0000 | 0.3491 | 0.6668 | — | 1.0 |
| | | B4 (*8i*) | 0.2598 | 0.3521 | -0.0444 | — | 1.0 |
| | *o*-B$_6$Se | Se1 (*4h*) | 0.0000 | 0.1701 | 0.8677 | — | 1.0 |
| | | B1 (*4h*) | 0.0000 | 0.1644 | 0.4812 | — | 1.0 |
| | | B2 (*8i*) | 0.8424 | 0.3672 | 0.3358 | — | 1.0 |
| | | B3 (*4h*) | 0.0000 | 0.3511 | 0.6608 | — | 1.0 |
| | | B4 (*8i*) | 0.2659 | 0.3535 | -0.0418 | — | 1.0 |
| CRYSTAL17 | *o*-B$_6$S | S1 (*4h*) | 0.0000 | 0.1851 | 0.8691 | — | 1.0 |
| | | B1 (*4h*) | 0.0000 | 0.1627 | 0.4827 | — | 1.0 |
| | | B2 (*8i*) | 0.8413 | 0.3618 | 0.3361 | — | 1.0 |
| | | B3 (*4h*) | 0.0000 | 0.3487 | 0.6666 | — | 1.0 |
| | | B4 (*8i*) | 0.2589 | 0.3515 | -0.0445 | — | 1.0 |
| | *o*-B$_6$Se | Se1 (*4h*) | 0.0000 | 0.1669 | 0.8667 | — | 1.0 |
| | | B1 (*4h*) | 0.0000 | 0.1637 | 0.4811 | — | 1.0 |
| | | B2 (*8i*) | 0.8418 | 0.3655 | 0.3358 | — | 1.0 |
| | | B3 (*4h*) | 0.0000 | 0.3505 | 0.6609 | — | 1.0 |
| | | B4 (*8i*) | 0.2641 | 0.3526 | -0.0421 | — | 1.0 |

[f] the atom site occupancies were fixed to 1.0;

**Table S2.** The frequencies of experimentally observed Raman bands ($\omega_0$) and Raman active modes predicted by VASP and CRYSTAL17 ($\omega_r^V$ and $\omega_r^C$), respectively. The overlapped band groups observed in experimental Raman spectra are presented by the corresponding frequencies regions; "—" signs indicate the absence of bands/band groups in the experimental Raman spectra.

### o-B$_6$S

| $\omega_0$, cm⁻¹ | $\omega_r^V$, cm⁻¹ | $\omega_r^C$, cm⁻¹ | Modes | $\omega_0$, cm⁻¹ | $\omega_r^V$, cm⁻¹ | $\omega_r^C$, cm⁻¹ | Modes | $\omega_0$, cm⁻¹ | $\omega_r^V$, cm⁻¹ | $\omega_r^C$, cm⁻¹ | Modes |
|---|---|---|---|---|---|---|---|---|---|---|---|
| 305.47 | 303.55 | 309.33 | B$_{1g}$ | 618.86 | 616.91 | 634.64 | A$_g$ | 780.02 | 781.84 | 802.09 | A$_g$ |
| 307-319 | 307.25 | 317.66 | A$_g$ | — | 640.41 | 660.56 | B$_{3g}$ | 809.17 | 802.65 | 824.64 | B$_{3g}$ |
|  | 308.65 | 318.39 | B$_{2g}$ |  | 655.42 | 673.72 | B$_{2g}$ | — | 835.46 | 852.28 | B$_{2g}$ |
| 334.46 | 331.40 | 340.95 | B$_{3g}$ | 673-706 | 682.21 | 698.92 | B$_{3g}$ |  | 878.89 | 896.82 | B$_{1g}$ |
| 347.05 | 338.70 | 350.08 | A$_g$ |  | 689.44 | 707.51 | A$_g$ | 939-1079 | 952.53 | 977.18 | A$_g$ |
| 373.25 | 363.94 | 378.07 | B$_{2g}$ |  | 694.10 | 712.08 | B$_{1g}$ |  | 950.51 | 978.60 | A$_g$ |
| 379.25 | 376.60 | 382.20 | B$_{1g}$ |  | 694.85 | 712.79 | B$_{2g}$ |  | 964.46 | 988.51 | B$_{3g}$ |
| 435.25 | 429.62 | 437.67 | B$_{3g}$ | 708-740 | 712.47 | 734.36 | A$_g$ |  | 962.88 | 989.83 | B$_{2g}$ |
| 466.54 | 460.38 | 470.99 | A$_g$ |  | 723.37 | 742.55 | B$_{1g}$ |  | 992.15 | 1022.98 | B$_{2g}$ |
| 470-490 | 464.96 | 475.70 | B$_{1g}$ |  | 730.64 | 751.44 | B$_{2g}$ |  | 992.92 | 1023.62 | A$_g$ |
|  | 483.37 | 488.10 | B$_{3g}$ | 742-772 | 745.54 | 766.73 | A$_g$ |  | 994.05 | 1025.61 | B$_{3g}$ |
| 496.15 | 488.92 | 498.44 | B$_{2g}$ |  | 751.81 | 768.42 | B$_{3g}$ |  | 1000.98 | 1030.45 | A$_g$ |
| 566.72 | 564.98 | 577.31 | B$_{1g}$ |  | 755.86 | 772.11 | B$_{1g}$ |  | 1015.79 | 1052.08 | B$_{2g}$ |
| 581.77 | 583.30 | 593.99 | B$_{2g}$ |  | 763.57 | 784.87 | B$_{2g}$ |  | 1016.45 | 1053.83 | A$_g$ |

### o-B$_6$Se

| $\omega_0$, cm⁻¹ | $\omega_r^V$, cm⁻¹ | $\omega_r^C$, cm⁻¹ | Modes | $\omega_0$, cm⁻¹ | $\omega_r^V$, cm⁻¹ | $\omega_r^C$, cm⁻¹ | Modes | $\omega_0$, cm⁻¹ | $\omega_r^V$, cm⁻¹ | $\omega_r^C$, cm⁻¹ | Modes |
|---|---|---|---|---|---|---|---|---|---|---|---|
| 192-201 | 191.69 | 197.87 | B$_{2g}$ | — | 612.38 | 632.73 | B$_{2g}$ | 824.84 | 824.75 | 843.56 | B$_{2g}$ |
|  | 192.52 | 198.57 | A$_g$ |  | 618.65 | 640.87 | B$_{3g}$ |  | 857.69 | 876.31 | B$_{1g}$ |
| 205.18 | 202.36 | 208.23 | B$_{1g}$ | 641-679 | 647.48 | 665.89 | A$_g$ | 921.87 | 907.53 | 940.63 | A$_g$ |
| 210.06 | 206.61 | 213.17 | B$_{3g}$ |  | 652.7696 | 669.76 | B$_{3g}$ | 933-1028 | 930.32 | 960.39 | B$_{3g}$ |
| 248.82 | 243.52 | 251.13 | A$_g$ |  | 661.47 | 680.76 | B$_{1g}$ |  | 946.73 | 965.96 | B$_{1g}$ |
| 260.55 | 255.44 | 263.67 | B$_{2g}$ |  | 668.24 | 686.26 | B$_{2g}$ |  | 939.27 | 968.32 | B$_{2g}$ |
| 347.69 | 344.68 | 349.24 | B$_{1g}$ | 682-784 | 686.33 | 704.41 | B$_{1g}$ |  | 931.71 | 973.02 | A$_g$ |
| 417.56 | 410.26 | 418.81 | B$_{3g}$ |  | 692.65 | 716.87 | A$_g$ |  | 961.14 | 984.44 | B$_{2g}$ |
| 429.24 | 422.95 | 433.59 | A$_g$ |  | 702.14 | 726.48 | B$_{2g}$ |  | 949.41 | 985.52 | B$_{3g}$ |
| 449.07 | 441.87 | 450.73 | B$_{1g}$ |  | 728.38 | 749.73 | B$_{3g}$ |  | 975.69 | 1001.11 | A$_g$ |
| 468-476 | 465.31 | 475.43 | B$_{3g}$ |  | 731.63 | 751.82 | A$_g$ |  | 984.30 | 1013.15 | A$_g$ |
|  | 468.69 | 475.86 | B$_{2g}$ |  | 742.69 | 762.78 | B$_{1g}$ |  | 990.05 | 1017.43 | B$_{2g}$ |
| 565.42 | 551.84 | 568.31 | B$_{1g}$ |  | 747.03 | 763.54 | B$_{2g}$ |  |  |  |  |
| 572.88 | 567.14 | 578.41 | B$_{2g}$ |  | 770.66 | 791.88 | A$_g$ |  |  |  |  |
| 594.62 | 591.37 | 610.71 | A$_g$ |  | 773.65 | 796.31 | B$_{3g}$ |  |  |  |  |